\documentclass[aps,pra,english,showpacs,twocolumn]{revtex4}
\usepackage{latexsym}
\usepackage{bm}
\usepackage{babel}
\usepackage[T1]{fontenc}
\usepackage[latin1]{inputenc}
\usepackage{mathrsfs}
\usepackage{amsmath, amssymb}
\usepackage{graphicx}
\usepackage{ae}

\begin{document}
\title{Influence of molecular symmetry on strong-field ionization:
Studies on ethylene, benzene, fluorobenzene, and
chlorofluorobenzene}
\author{Thomas K. Kjeldsen}
\author{Christer Z. Bisgaard}
\author{Lars Bojer Madsen}
\affiliation{Department of Physics and Astronomy,
  University of Aarhus, 8000 \AA rhus C, Denmark}
\author{Henrik Stapelfeldt}
\affiliation{Department of Chemistry, University of
  Aarhus, 8000 \AA rhus C, Denmark}

\begin{abstract}
Using the molecular strong-field approximation we consider the
effects of molecular symmetry on the ionization of molecules by a
strong, linearly polarized laser pulse. Electron angular
distributions and total ionization yields are calculated as a
function of the relative orientation between the molecule and the
laser polarization. Our studies focus on ethylene (C$_2$H$_4$),
benzene (C$_6$H$_6$), fluorobenzene (C$_6$H$_5$F), and ortho
chlorofluorobenzene (1,2 C$_6$H$_4$ClF), the molecules
representing four different point groups. The results are compared
with experiments, when available, and with the molecular tunneling
theory appropriately extended to non-linear polyatomic molecules.
Our investigations show that the orientational dependence of
ionization yields is primarily determined by the nodal surface
structure of the molecular orbitals.

\end{abstract}

\maketitle

\section{Introduction}
\label{sec:intro}

Ionization is an ubiquitous process in the interaction between
atoms and molecules and intense laser fields. The ejected electron
propagates in the combined laser and Coulomb field, and the laser
field may steer  back the electron and force it to recollide with
the parent ion. In this way, the release of an electron into the
laser-dressed continuum can initiate strong-field processes, such
as above threshold ionization, multiple ionization, and high
harmonic generation (see, e.g., Ref.~\cite{brabec:2000:rmp} and
references therein). Clearly, the laser field only drives the
electron in the direction of the laser polarization vector and,
consequently, rescattering dynamics is typically only present for
linearly polarized laser fields. The possibility of rescattering
depends not only on the parameters of the laser. Also the
scattering angle with which the electron leaves the parent ion
plays a role. For most atomic ground states, strong-field
ionization leads to an outgoing electronic current in the
direction of the linear polarization of the laser, and hence a
recollision is possible after the field has changed its direction.
For molecules, however, it happens, that the molecular symmetry
imposes one or more nodal planes in the ground state wave
function. In such planes the ejection of photoelectrons may be
strongly suppressed \cite{kjeldsen03}, causing recollisionally
induced double ionization not to peak for linearly but for
elliptically polarized laser light \cite{bhardwaj01}. It is
therefore clear that  a good understanding of the dependence of
the initial ionization process upon laser parameters and upon
parameters characteristic for the system, such as symmetry, is
crucial for the further development and possible practical
utilization of strong-field phenomena.

A large number of studies has established that the laser
parameters determining strong-field ionization are the wavelength,
the intensity, and the polarization
\cite{brabec:2000:rmp,markevitch04,dewitt99,lezius01,lezius02}.
We note in passing that in the case of few-cycle laser pulses, the
ionization rate and the photoelectron angular distribution also
depends critically on the phase of the carrier frequency with
respect to the envelope \cite{paulus01}. For atoms the most
important system parameter is the binding energy of the initial
state. In fact, the simple ADK tunneling model \cite{ADK},
employing only the laser intensity and the atomic binding energy
as input parameters, provides for many atomic species a very good
description of measured total ionization rates. Also, the theory
referred to as the strong-field approximation (SFA)
\cite{Keldysh,Faisal,reiss80,krainov97,Larochelle} has been
successful in modeling atomic ionization including the angular
distribution of the ejected photoelectrons. In this theory, the
ionization takes place from a field-free ground state to a
laser-dressed continuum final state via the interaction with the
lowest-order laser-atom operator. In the final state, the Coulomb
interaction between the electron and the parent ion is neglected
and the electron is described by the state of a free electron in
the laser field -- the so-called Volkov state (see
Ref.~\cite{Madsen05} for a very recent discussion).

For molecules a few recent studies have shown that there is an
additional system parameter that also strongly influences
strong-field ionization namely the symmetry of the initial
electronic state
\cite{corkum03,eremina04,alnaser04,muthbohm00,muthbohm01,lin02,lin03,kjeldsen03,kjeldsen04}.
Modifications of both the ADK tunneling theory \cite{lin02} and
the SFA \cite{muthbohm00} have been introduced to account for the
molecular symmetry. Hereby the two models have provided results in
reasonably agreement with measured ionization rates of molecules.
In the molecular ADK (MO-ADK) theory \cite{lin02} one expands the
asymptotic molecular ground state wave function in a one-center
basis and applies the atomic tunneling theory to each partial 
wave. The atomic SFA is straightforwardly generalized to the
molecular SFA (MO-SFA) by using molecular wave functions. A very
appealing feature of MO-SFA is that the expansion of the molecular
wave function in linear combination of atomic orbitals (LCAO),
allows an interpretation of the spectrum of photoelectrons in
terms of interference from the multi-center molecule; an
interpretation which  in the context of single-photon ionization
was offered almost 40 years ago \cite{Cohen66}.

In our previous work \cite{kjeldsen03} we used the molecular
strong-field approximation (MO-SFA) in the velocity gauge
formulation to illustrate the strong dependence of the
photoelectron angular distribution and the total ionization rates
of ethylene and benzene on (i) the orientation between the
molecule and the laser polarization and on (ii) the symmetry of
the electronic wave function. The purpose of the present paper is
to give a more thorough discussion of the influence of molecular
symmetry on strong-field ionization. In particular, we study four
different molecules representing four different classes of
symmetry (point groups). We show how a reduction of the molecular
symmetry lifts the degeneracy of molecular orbitals and leads to
different ionization channels with distinct features.

More recently, we found that the MO-SFA in the velocity gauge
formulation failed to explain orientational effects in ionization
of N$_2$. However, using the length-gauge to describe the
molecule-field interaction, results in good agreement with
experiments~\cite{corkum03} were obtained~\cite{kjeldsen04}.
Consequently, in the present work we found it necessary to employ
both gauges and check if the results obtained are consistent. In
addition, we calculate ionization rates by the MO-ADK tunneling
model and compare with the results from the MO-SFA.

The paper is organized as follows. In Sec.~\ref{sec:structure},
we present a
justification for using a Hartree-Fock calculation to represent
the electronic ground states of the molecules under study. In
Sec.~\ref{sec:theory}, we recall the theory of the MO-SFA and the MO-ADK
tunneling models. In particular, we present the generalization of
the MO-ADK theory~\cite{lin02} to the case of non-linear
polyatomic molecules. In Sec.~\ref{sec:calc}, we provide details of the
calculations  and discuss the choice of basis functions, the
choice of nuclear coordinates, and the method for obtaining the
ionization rates. In Sec.~\ref{sec:results}, we present the numerical results
comprising a comparison between the length and the velocity gauge.
We argue that the length gauge gives the better results and use it
together with the MO-ADK model to calculate ionization rates for
each of the four molecules. In Sec.~\ref{sec:conclusions}, we conclude and provide an
outlook for the future.

\section{Electronic structure: Hartree-Fock or beyond}
\label{sec:structure}

In some molecules the inclusion of electron correlations is
necessary in order to predict even qualitatively correct
properties. For example, in N$_2$ the ordering of the orbitals
obtained from a simple Hartree-Fock (HF) calculation is not
consistent with the observed single-photon photoelectron spectrum:
The calculated highest occupied molecular orbital (HOMO) of N$_2$
is a $\pi_u$ orbital while the HOMO inferred from experiments is a
$\sigma_g$ orbital~\cite{data}. When including electron
correlations in the calculations, the correct ordering and
ionization energies are obtained \cite{bartlett88}. Another
example where the effects of electron--electron correlation are
important is F$_2$ which is not even bound in a HF treatment.
It is therefore not surprising that one--electron models have
failed in the attempt of explaining strong-field ionization of
F$_2$ \cite{muthbohm00,lin02,Benis04}.

In this work, we consider ionization of ethylene and substituted
benzenes. For these systems electron correlation may be safely
neglected in the description of the initial state, since
photoionization spectra in the valence region of these molecules
consist of clearly separated bands which are unambiguously
assigned and are in accordance with Hartree-Fock predictions
\cite{holland97,baltzer97}. In these systems, each electron
occupies an orbital obtained from a HF calculation and it is an
accurate approximation to describe ionization by the removal of an
electron from an occupied orbital while leaving the ion in an
unrelaxed state. With these approximations, the many--electron
problem translates into the much simpler single--electron problem.
Only ionization from the highest or next highest occupied 
molecular orbital (referred to as HOMO--1) is considered. 
For ethylene and benzene the
separations between the first and second ionization potential are
$2.3\, \mbox{eV}$ and $2.2\, \mbox{eV}$, respectively. In
Ref.~\cite{jaron04} it was shown that such large energy
differences led to a much lower ionization probability from the HOMO--1
than from the HOMO. The separation between
the first and second ionization potential is, however, much lower
for the substituted benzenes, approximately $0.4\, \mbox{eV}$.
Accordingly, we will consider ionization from the HOMO only for
ethylene and the degenerate HOMO of benzene while for the
substituted benzenes we will additionally include ionization from the HOMO--1.
Ionization from all other orbitals is neglected.

\section{Theory}
\label{sec:theory}

In this section, we describe the theories that we apply for the
calculation of the molecular strong-field ionization. Throughout
this work, we fix the nuclei at their equilibrium positions as
determined with the HF theory.

\subsection{Molecular strong-field approximation}
\label{sec:sfatheory}
The molecular strong-field approximation (MO-SFA)
\cite{muthbohm00,kjeldsen03,kjeldsen04} is a
generalization of the atomic strong-field approximation
\cite{Keldysh,Faisal,reiss80}.
For a linearly polarized electric field ${\bm F}(t) =
{\bm F}_0 \sin{\left( \omega t \right)}$ with associated vector potential ${\bm
  A}(t) = {\bm A}_0 \cos{\left( \omega t \right)}$, $A_0 = F_0 / \omega$,
the laser electron interaction is represented by [atomic units
($\hbar = m_e = e = a_0= 1$) are used throughout]
\begin{equation}
    \label{eq:intlen}
    V_F^{(\text{LG})}(t) = {\bm r} \cdot {\bm F}(t),
\end{equation}
in the length gauge (LG), and by
\begin{equation}
    \label{eq:intvel}
    V_F^{(\text{VG})}(t) = {\bm A}(t) \cdot {\bm p} + \frac{{\bm A}^2(t)}{2},
\end{equation}
in the velocity gauge (VG).

In either gauge, we utilize the $2\pi /\omega= T$ periodicity of
the field to express the angular differential, $d W/d\hat{{\bm
q}}$, and total, $W$, ionization rates as sums of $n$-photon
absorptions \cite{Gribakin97,kjeldsen04}
\begin{eqnarray}
    \label{eq:angdiffrate}
    \frac{dW}{d\hat{{\bm q}}} &=& 2 \pi
    \sum_{n=n_0}^\infty |A_{{\bm q}n}|^2 q_n,\\
    \label{eq:totrate}
    W &=& 2 \pi \sum_{n=n_0}^\infty \int |A_{{\bm q}n}|^2 q_n d\hat{{\bm q}},
\end{eqnarray}
where the transition amplitudes
$A_{{\bm q}n}= 1/T \int_0^T\langle \Psi_V({\bm r},t)|V^{(c)}_F(t)|\Psi({\bm
r},t)\rangle dt$ ($c=\{\text{LG,VG}\}$)
are evaluated at the momentum $q_n = \sqrt{2(n \omega - E_b -
  U_p)}$, with $E_b$ the binding energy of the initially bound electron
and $U_p$ the ponderomotive potential. Energy conservation also determines
the minimum number of photons needed to reach the continuum, i.e., the
lower limit $n_0$ of the sum. In the expression for $A_{{\bm q}n}$, $\Psi_V$
is a $(2 \pi)^{-3/2}$ normalized Volkov wave function and $\Psi({\bm
  r},t)$ the wave function for the electron in the combined field of
the molecule and the laser. Introducing the strong-field
approximation, i.e., $ \Psi({\bm r},t) \simeq \Psi_0({\bm r},t) =
e^{i E_b t} \Phi_0({\bm r})$, with $\Phi_0$   the solution of the
time-independent field--free HF equation, we find the length and
velocity gauge amplitudes \cite{Gribakin97,kjeldsen04}
\begin{widetext}
\begin{equation}
  \label{eq:ltransamp}
  A_{{\bm q}n}^{(\text{LG})} = \frac{1}{T} \int_0^T dt \left(-E_b -
  \frac{Q^2_n(t)}{2} \right)
  \tilde{\Phi}_0({\bm Q}_n(t))
  \exp i\left(n \omega t+{\bm q}_n \cdot
  \boldsymbol{\alpha}_0\sin(\omega
  t)+\frac{U_p}{2 \omega}\sin(2\omega t)\right),
\end{equation}
\begin{equation}
  \label{eq:vtransamp}
  A_{{\bm q}n}^{(\text{VG})} = \left(-E_b -
  \frac{q_n^2}{2} \right)
  \tilde{\Phi}_0({\bm q}_n)
  \frac{1}{T} \int_0^T dt \exp{ i\left( n \omega t+{\bm q}_n\cdot
  \boldsymbol{\alpha}_0\sin(\omega t)+\frac{U_p}{2 \omega}\sin(2\omega
  t)\right)}.
\end{equation}
\end{widetext}
Here ${\bm Q}_n(t) = {\bm q}_n+{\bm A}(t)$ denotes the time--dependent
momentum, $\boldsymbol{\alpha}_0 ={\bm A}_0/\omega$ denotes the quiver
radius, and
$\tilde{\Phi}_0 ({\bm q})= (2\pi)^{-3/2} \int
d {\bm r} e^{-i {\bm q} \cdot {\bm r}} \Phi_0({\bm r})$ the
momentum wave function of the initially bound electron. The velocity
gauge transition amplitude of Eq.~(\ref{eq:vtransamp}) may be simplified by
replacing the time integral with a generalized Bessel function
$J_n(u,v)$ \cite{reiss80} and using energy conservation
\begin{equation}
    \label{eq:vtransamp2}
    A_{{\bm q}n}^{(\text{VG})} = \left(U_p - n \omega \right)
    \tilde{\Phi}_0({\bm q}_n) J_{-n} \left( {\bm q}_n\cdot
    \boldsymbol{\alpha}_0,
    \frac{U_p}{2 \omega} \right).
\end{equation}
Applying Eq.~(\ref{eq:vtransamp2}) to the differential rate
of Eq.~(\ref{eq:angdiffrate}) is equivalent to the expression used in
Ref.~\cite{muthbohm00}.
We note that Eq.~(\ref{eq:ltransamp})
simplifies to Eq.~(\ref{eq:vtransamp}) when neglecting ${\bm A}(t)$
compared to ${\bm q}_n$, i.e., when ${\bm Q}_n(t) \approx {\bm q}_n$.

We construct the molecular one-electron wave functions as linear
combinations of basis functions $\phi_{ij}$ based on each atom
(LCAO-MO), located at the positions $\bm R_i$. The wave function
can now be written as
\begin{equation}
 \Phi_0(\boldsymbol{r},\{\boldsymbol{R}_i \} )= \!\!
 \sum_{i=1}^N \sum_{j=1}^{j_{max}^{(i)}}\!\!\!
a_{ij} \phi_{ij}(\boldsymbol{r} - \boldsymbol{R}_i) ,
\label{eqn:LCAO}
\end{equation}
where $i$ runs over the $N$ atoms in the molecule and $j$ runs
over the basis functions on each atom $\phi_{ij}$. The Fourier
transform of the wave function is then 
\begin{equation}
  \tilde{\Phi}_0({\bm q}_n) = \sum_{i=1}^N \sum_{j=1}^{j_{max}^{(i)}} a_{ij}\tilde{\phi}_{ij}({\bm q}_n)
  e^{-i\boldsymbol{q}_n\cdot\boldsymbol{R}_i}, 
\label{eqn:fourier}
\end{equation}
with $\tilde{\phi}_{ij}$ representing $\phi_{ij}$ in momentum space.

\subsection{Molecular tunneling theory}
\label{sec:adktheory}

The molecular ADK (MO-ADK) tunneling theory \cite{lin02,lin03} is
a generalization of the atomic ADK theory \cite{ADK}. In previous
works \cite{lin02,lin03,kjeldsen04} the MO-ADK theory was applied
to diatomic molecules. Here the theory is generalized to cover
also nonlinear polyatomic molecules.

The tunneling theory relies on the assumption that at any given instants of
time the molecule will respond to the external laser field as if it were a
static electric field.
The rate of ionization in the oscillating field may thus be determined
by the time-averaged static rates. Whether this
approach is reasonable or not depends on the value of the Keldysh parameter
$\gamma = \sqrt{2E_b}\frac{\omega}{F_0}$ \cite{Keldysh} with $\gamma \ll 1$
in the tunneling regime.

The formulation of a
tunneling theory in the case of molecules is complicated compared
to the atomic case by two related features. Firstly, the presence
of multiple nuclei within the molecule breaks the spherical
symmetry of the field--free system and necessitates a description
of the molecular wave function in terms of a superposition of
partial waves. Secondly, the Euler angles describing the
orientation between the laboratory fixed frame (with a $Z$--axis
determined by the linear polarization vector of the external
field) and the molecular body fixed frame have to be specified.
Note that we use $(X,Y,Z)$ to label the laboratory fixed frame of
reference and $(x,y,z)$ to label the body--fixed frame of reference.
The tunneling rate of molecules can be determined once the
field--free asymptotic wave function is known. In a body--fixed frame,
labeled by superscript $B$, this
function must follow the asymptotic Coulomb form
\begin{equation}
  \label{eqn:bodywf}
  \Phi^B_0(\bm{r}) \sim r^{Z_\text{ion}/\kappa -1} e^{- \kappa r} \sum_{l,m}
C_{lm} Y_{lm}(\hat{\bm{r}}),
\end{equation}
where $Z_\text{ion}$ is the charge of the residual ion and
$\kappa$ is related to the binding energy, $\kappa = \sqrt{2
E_b}$. 
For the tunneling process in the DC case, we assume the electric field to
point in the positive $Z$--direction, corresponding to a situation where
tunneling would occur in the negative $Z$--direction. Consequently, we need
to express the asymptotic form of the molecular wave function in that
direction.
If the
body--fixed frame is rotated by the Euler angles ($\phi, \theta,
\chi$) with respect to the laboratory fixed system, the asymptotic
form in the laboratory fixed system of the field--free molecular
ground state wave function is expressed through the rotation
operator $\hat{D}(\phi, \theta, \chi)$ as $\Phi^L_0(\bm{r}) =
\hat{D}(\phi, \theta, \chi) \Phi^B_0(\bm{r})$, where we have used
the superscript $L$ to denote the laboratory fixed system.
Equation~(\ref{eqn:bodywf}) then leads to the expression
\begin{equation}
  \label{eqn:labwf}
  \Phi^L_0(\bm{r}) \sim r^{Z_\text{ion}/\kappa -1} e^{- \kappa r} \sum_{l,m}
C_{lm} \sum_{m'} \mathscr{D}_{m'm}^{(l)} (\phi, \theta, \chi) Y_{lm'}
(\hat{\bm{r}}),
\end{equation}
where $\mathscr{D}_{m'm}^{(l)} (\phi, \theta, \chi)$ is a Wigner
rotation function \cite{brink,zare}. In Eq.~(\ref{eqn:labwf}) the
sum over $m'$ and the corresponding Wigner functions describes the
rotation between the coordinate systems and the sum over $l,m$ is
a signature of the breaking of the spherical symmetry by the
molecular system. For linear molecules the projection $m$ of the
electronic angular momentum onto the body--fixed axis is a good
quantum number and, hence, for such systems there would be no
summation over $m$ in Eqs.~(\ref{eqn:bodywf}) and
(\ref{eqn:labwf}) \cite{lin02}. For later convenience, we note that
$\phi$ and $\chi$ represent rotations around the space--fixed $Z$--axis and
the body fixed $z$--axis, respectively, while $\theta$ represents the angle
between the $Z$ and $z$ axes.

From the asymptotic form of Eq.~(\ref{eqn:labwf}), the total ionization rate in a static (DC)
field in the positive $Z$ direction is calculated as in the atomic
case \cite{Smirnov66,Perelomov66,Bisgaard04}, and the result is
\begin{eqnarray}
  \label{eqn:statrate}
  W_\text{stat}(F_0) =
  \sum_{m'} \frac{|B(m')|^2}{2^{|m'| } |m'|!
\kappa^{2z/\kappa-1}}\\ \nonumber \times \left( \frac{2\kappa
^3}{F_0}\right)^{2Z/\kappa - |m'| - 1} \exp \left( -\frac{2}{3}
\frac{\kappa^3}{F_0}\right),
\end{eqnarray}
where
\begin{eqnarray}
  \label{eq:bm}
  B(m') = \sum_{l,m} (-1)^{(| m' | + m')/2} \sqrt{\frac{(2l+1)(l + | m'|)!}
{2(l-|m'|)!}}\\
\nonumber \times C_{lm} \mathscr{D}_{m' m}^{(l)} (\phi, \theta, \chi).
\end{eqnarray}

In a slowly varying field, the ionization rate is found by averaging the DC
rate over an optical cycle
\begin{equation}
    \label{eq:tunnelrate}
    W = \frac{1}{2\pi}\int_{0}^{2\pi} W_\text{stat}\left(F_0\cos(\omega t) \right)
    d(\omega t).
\end{equation}
The DC rate is given by Eq.~(\ref{eqn:statrate}) when the field is
oriented in the positive $Z$ direction, $|\pi - \omega t|\ge \pi/2$,
corresponding to tunneling in the negative $Z$--direction. When
the field points in the negative $Z$ direction, $\pi/2 \le \omega t \le
3\pi/2$, corresponding to the possibility for tunneling in the positive
$Z$--direction, the geometry is equivalent to a field pointing in the positive
$Z$ direction but with an inverted molecule.
We account for this by applying the parity operator $\mathscr{P}$ on
the wave function $\Phi^B_0$
\begin{equation}
  \mathscr{P} \Phi^B_0(\bm r) \sim r^{Z_\text{ion}/\kappa-1}e^{-\kappa
  r}\sum_{l,m}C_{lm}(-1)^l Y_{lm}(\hat{\bm r}),
\end{equation}
and we see by comparing with Eq.~(\ref{eqn:bodywf}) that this
simply corresponds to the substitution $C_{lm}\rightarrow (-1)^l
C_{lm}$ in Eq. (\ref{eq:bm}). Note that parity eigenstates, corresponding
to inversion symmetric molecules,
contain only even or odd $l$ states. When the field direction is changed
Eq.~(\ref{eq:bm}) will either remain invariant or change sign
and the DC rate is thereby left invariant.
Contrary, the DC rate will not be
invariant to field inversion for states which are not parity eigenstates.
This is also to be expected since the wave functions in the tunneling
regions in the positive and negative $Z$ direction will be different.
Under the assumption $\kappa^3/F_0 \ll 1$ the integral in 
Eq.~(\ref{eq:tunnelrate}) may be approximated by
\begin{equation}
  \label{eq:tunnelrate2}
  W = \sqrt{\frac{3F_0}{\pi \kappa^3}}\frac{W^+_\text{stat}(F_0) +
  W^-_\text{stat}(F_0)}{2},
\end{equation}
where $W^\pm_\text{stat}(F_0)$ are the DC rates for the positive and negative
field directions with respect to the $Z$ direction.

\section{Calculations}
\label{sec:calc}

\subsection{Basis set}
\label{sec:basis}

The Hartree-Fock wave functions for ethylene, benzene, fluorobenzene and
{\it o}--chlorofluorobenzene are calculated using GAMESS
\cite{gamess} with a standard valence triple zeta basis set \cite{dunning71}.
This basis set contains $s$ and $p$ orbitals for carbon, fluorine and
chlorine while the hydrogen basis consists only of $s$
orbitals. Since the tunneling rates and the length gauge MO-SFA rates both rely
on the asymptotic form of the wave function a precise description is needed
in this region. This is obtained by adding an extra diffuse $s$ and $p$
basis function.
The orbitals of the molecules considered are odd under
reflection in their molecular planes. Thus the LCAO expansion,
Eq.~(\ref{eqn:LCAO}), can only include $p$ orbitals orthogonal to the plane.
For example, if we
define the $yz$ plane to coincide with the molecular plane, the relevant
$p$ orbitals are the $p_x$ orbitals.

\subsection{Expansion and asymptotic coefficients}
\label{sec:expansion} 
\begin{figure}
      \begin{center}
      \includegraphics[width=0.8\columnwidth]{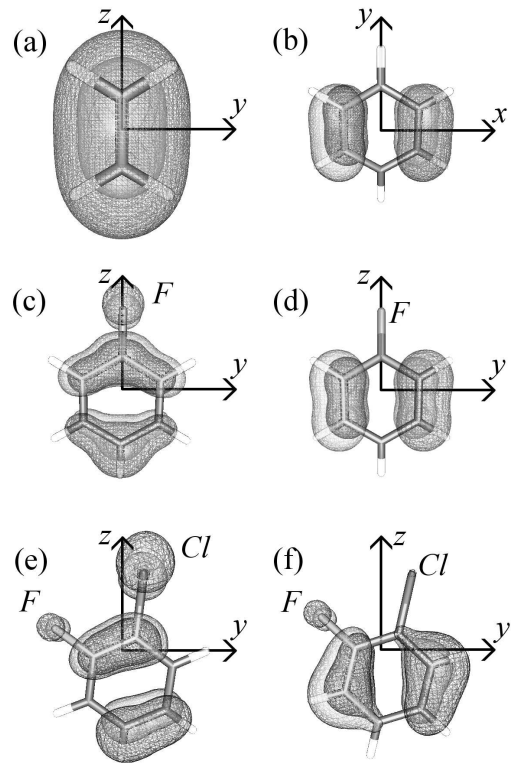}
    \end{center}
    \caption{Nuclear geometries and isocontours for the active orbitals of (a)
    ethylene, (b) benzene, (c) and (d) HOMO and HOMO--1 for fluorobenzene
    respectively and (e) and (f) HOMO and HOMO--1 for
    1,2-chlorofluorobenzene respectively. 
    The origins of the coordinate systems are in
    the center--of--mass. For later convenience we see that all the orbitals
    contain nodal planes in the molecular plane. The HOMOs of fluorobenzene
    and 1,2-chlorofluorobenzene, (d) and (f), contain nodal surfaces which
    are approximately parallel to the $xy$ plane. The HOMO--1 of
    fluorobenzene contains a nodal plane in the $xz$ plane while the HOMO--1
    of 1,2-chlorofluorobenzene contains a nodal suface which is almost in
    the $xz$ plane. The relation between the body--fixed axes shown in the figures
    and the laboratory--fixed axes is discussed in Sec.~\ref{sec:adktheory}.}
    \label{fig:structure}
\end{figure}

The calculation of the initial molecular
ground state wave function is performed at the equilibrium
geometry predicted by Hartree--Fock theory. Bond lengths obtained
in this way are typically within few pm from experimental data
\cite{helgaker00}. The nuclear geometries are shown in
Fig.~\ref{fig:structure}.

Note that we define the body--fixed $z$ axis to be
perpendicular to the molecular plane for benzene while it lies in the
molecular plane for the remaining molecules.
Since we want to investigate  the
influence of molecular alignment on the ionization rates,
we have chosen our axes to coincide with the principal axes for
each molecule.
Aligned molecules
are generated as superpositions of rotational eigenstates which are most
conveniently described in this coordinate system \cite{eckart}.
Our definitions of coordinates are in
accordance with the usual conventions except for chlorofluorobenzene.

The asymptotic form of the orbitals, Eq.~(\ref{eqn:bodywf}), is found by
expanding in spherical harmonics
\begin{equation}
  \Phi_0(\bm r) = \sum_{l,m}F_{lm}(r)Y_{lm}(\hat{\bm r}),
\end{equation}
leading to
\begin{equation}
  \label{eqn:flm}
  F_{lm}(r) = \int d\Omega Y^*_{lm}(\hat{\bm r}) \Phi_0(\bm r).
\end{equation}
These functions are then matched to the form $F_{lm}(r) \sim C_{lm}r^{Z_\text{ion}/\kappa - 1}
e^{-\kappa r}$ in the asymptotic region whereby the coefficients
$C_{lm}$ are determined. The values of $C_{lm}$ for the molecules
investigated in this work are given in Table~\ref{tab:clm}. All the orbitals
are odd eigenstates of the $\sigma_{yz}$ reflection operation and they are
additionally chosen purely real. The
projection on the spherical harmonics, Eq.~(\ref{eqn:flm}),  will 
then be either purely real ($m$ odd) or purely imaginary ($m$ even).
Another consequence of the $yz$ reflection antisymmetry is 
$F_{l,-m}(r) = -F_{l,m}(r)$. 
In ethylene, benzene and fluorobenzene our orbitals
are additionally be eigenstates of the $\sigma_{xz}$ reflection operation.
If the wave function is an even (odd) eigenstate of the $\sigma_{xz}$ reflection 
operation then only the real, $m$ odd, (imaginary, $m$ even) radial
functions will occur.

\begin{table}
\caption{Asymptotic coefficients $C_{lm}$.}
\begin{ruledtabular}
  \begin{tabular}{|l|l|c|c|c|c|c|c|}
    \label{tab:clm}
    $l$&  $m$  &C$_2$H$_4$&C$_6$H$_6$
    &\multicolumn{2}{c|}{C$_6$H$_5$F}&\multicolumn{2}{c|}{1,2 C$_6$H$_4$ClF} \\
    &       &          &          & HOMO     & HOMO--1   & HOMO     & HOMO--1 \\
\hline
1   &$\pm 1$&$\pm 1.10$&          &$\pm 0.49$&          &$\pm 1.27$&$\pm 0.08$\\
\hline
2   &$\pm 2$&          &          &          &$\mp 1.70i$&$\mp 0.39i$&$\mp 1.39i$\\
    &$\pm 1$&          &$\pm 1.36$&$\mp 1.70$&          &$\mp 0.95$ &$\pm 0.19$\\
\hline
3   &$\pm 3$&          &          &$\pm 0.13$&          &$\mp 0.28$&$\mp 0.54$\\
    &$\pm 2$&          &          &          &$\pm 1.22i$&$\pm 0.39i$&$\pm 1.51i$\\
    &$\pm 1$&$\pm 0.22$&          &$\pm 0.89$&          &$\pm 1.83$&$\mp 0.71$\\
\hline
4   &$\pm 4$&          &          &          &$\pm 0.35i$&$\pm 0.10i$&$\pm 0.41i$\\
    &$\pm 3$&          &          &$\pm 0.24$&          &$\pm 0.13$&$\pm 0.16$\\
    &$\pm 2$&          &          &          &$\mp 0.68i$&$\mp 0.63i$&$\mp 0.75i$\\
    &$\pm 1$&          &$\mp 0.32$&$\pm 0.31$&          &$\mp 0.53$&$\pm 0.26$\\
\hline
5   &$\pm 5$&          &          &          &          &$\pm 0.03$&$\pm 0.11$\\
    &$\pm 4$&          &          &          &$\mp 0.25i$&$\mp 0.10i$&$\mp 0.30i$\\
    &$\pm 3$&          &          &$\mp 0.11$&          &$\mp 0.21$&$\mp 0.01$\\
    &$\pm 2$&          &          &          &$\pm 0.33i$&$\pm 0.06i$&$\pm 0.30i$\\
    &$\pm 1$&          &          &$\pm 0.23$&          &$\pm 0.82$&$\mp 0.25$\\
\hline
6   &$\pm 4$&          &          &          &$\pm 0.16i$&$\pm 0.10i$&$\pm 0.18i$\\
    &$\pm 2$&          &          &          &$\mp 0.12i$&$\mp 0.22i$&$\mp 0.10i$\\
    &$\pm 1$&          &          &          &          &$\mp 0.17$&$\pm 0.07$\\
\end{tabular}
\end{ruledtabular}
\end{table}
Since the basis functions we use are Gaussian and the correct
asymptotic form to which we fit, is exponential, we cannot expect
a precise fit all the way to infinity. Accordingly, we have
adjusted the range of asymptotic fitting to start well outside the
nuclear positions and is limited such that the radial functions
$F_{lm}(r)$ of Eq.~(\ref{eqn:flm}) still follow the exponential
behavior. In Ref.~\cite{kjeldsen04}, we obtained accurate results
by this procedure.

\subsection{Simplifications in the length gauge}\label{sec:lengthcalculation}

In the length gauge MO-SFA the transition amplitude, will only
depend on the asymptotic behavior of the coordinate space wave
function \cite{Gribakin97,kjeldsen04}.

When we have determined the angular coefficients $C_{lm}$ we may
simply use the Fourier transform of Eq.~(\ref{eqn:labwf}) in the
length gauge transition amplitude of Eq.~(\ref{eq:ltransamp}). The
Fourier transform of Eq.~(\ref{eqn:labwf}) is
\begin{equation}
  \label{ftr-phi0}
  \tilde{\Phi}_0(\bm q) =
  \sum_{l}\tilde{f}_l(q)\sum_{m,m'}C_{lm}\mathscr{D}^{(l)}_{m'm}(\phi,\theta,\chi)Y_{lm'}(\hat{\bm
  q}),
\end{equation}
where the radial momentum--space functions $\tilde{f}_l(q)$ are
\begin{eqnarray}
  \label{ftr-fl}
  \tilde{f}_l(q) &= &2\pi^{3/2}\left(-\frac{iq}{2}\right)^l
  \frac{\Gamma(l+\frac{Z}{\kappa}+2)}{\kappa^{l+Z/\kappa+2}\Gamma(l+\frac{3}{2})}\\
  \nonumber
&\times&  {}_2F_1\left(
\frac{l+\frac{Z}{\kappa}+2}{2},\frac{l+\frac{Z}{\kappa}+3}{2};
l+\frac{3}{2};
  -\frac{q^2}{\kappa^2} \right),
\end{eqnarray}
with $_2F_1(a,b;c;z)$ Gauss' hypergeometric function \cite{stegun}.
Equation (\ref{eq:ltransamp}) can now be rewritten as
\begin{eqnarray} \label{Aqn2}
  A_{\bm{q}n}^{(\text{LG})} &=&\sum_{l,m,m'} \mathscr{D}^{(l)}_{m'm}(\phi,\theta,\chi)C_{lm}\\
  \nonumber
  &\times&
  \frac{1}{T} \int_{0}^{T} dt (-E_b -
  \frac{Q^2_n(t)}{2}) \tilde{f}_l(Q_n(t))Y_{lm'}(\hat{\bm Q }_n(t))\\
\nonumber &\times& \exp i \left(n \omega t + \bm{q}_n \cdot
\bm{\alpha}_0 \sin ( \omega t) + \frac{U_p}{2 \omega} \sin (2
\omega t)\right),
\end{eqnarray}
with $\bm Q_n(t) =
\bm q_n + \bm A(t)$. This formula is particularly useful when we
want to consider many different molecular orientations, since the
integrals in Eq.~(\ref{Aqn2}) are independent of molecular
orientation. Firstly, we need to calculate all the integrals once
for each $l$ and $m'$ at some lab fixed momentum $\bm q_n$. Then
we can easily get the transition amplitude at the same lab fixed
momentum for all orientations $(\phi,\theta,\chi)$ by
multiplication with the appropriate Wigner functions.
We have checked this method against calculations with the fully
numeric Fourier transforms applied in Eq.~(\ref{eq:ltransamp}) and
we find agreement within $\lesssim 20\%$ in the final rates.

In closing this section, we note that with the above simplifications of the
length gauge MO--SFA, the only non--trivial dependency on electronic
structure of the MO--SFA LG and the MO--ADK theory is through the
$C_{lm}$ expansion coefficients.

\subsection{Ion signal}
\label{sec:signal}
The ionization rates in Eqs.~(\ref{eq:totrate}) and (\ref{eq:tunnelrate}) depend on
the instantaneous amplitude
of the field and the molecular orientation described by the Euler angles
($\phi, \theta, \chi$). We only consider linearly polarized light and hence
the results are independent of $\phi$, the angle of rotation around the
polarization vector.
For a Gaussian laser beam with a Gaussian temporal profile with FWHM $\tau$,
the amplitude of the vector potential is
\begin{equation}
  A({\cal R,Z},t) = \frac{\sqrt{I_0}}{\omega} \frac{ w_0}{w({\cal Z})} e^{-
  {\cal R}^2/w({\cal Z})^2} \exp \left(-\frac{2 \ln 2 t^2}{\tau^2} \right),
\end{equation}
where $({\cal R,Z})$ are the cylindrical coordinates, $I_0$ the peak intensity,
$w_0$ is the
spot size and $w({\cal Z})=w_0 \sqrt{1+{\cal Z}^2/{\cal Z}_R^2}$ where
${\cal Z}_R = \pi w_0^2/\lambda$ is the Rayleigh length and $\lambda$
the wavelength.

The rate equations for the ionization probability of a molecule oriented
according to the Euler angles ($\theta, \chi$) and
located at $({\cal R,Z})$ in the laser focus are
\begin{eqnarray}
    \label{eqn:rateeq}
    \frac{dp_0}{dt} &=& - W\left( \boldsymbol A({\cal R,Z},t), \theta, \chi \right) p_0\\
    \frac{dp_1}{dt} &=&  W\left( \boldsymbol A({\cal R,Z},t), \theta,
    \chi \right) p_0,
\end{eqnarray}
where
$p_0$ and $p_1$ denote the probabilities of having a neutral or
an ion, and ionization to higher charge states is neglected. By
the end of the pulse, the solution to Eq.~(\ref{eqn:rateeq}) is given by
\begin{equation}
    \label{eqn:rate_sol}
    p_1({\cal R,Z}, \theta,\chi) = 1 - \exp\left(- \int_{-\infty}^\infty W\left( \boldsymbol
    A ({\cal R,Z}, t'), \theta, \chi \right) dt' \right).
\end{equation}
The orientational dependent number of ionized molecules $N_1$ is found by
integrating Eq.~(\ref{eqn:rate_sol}) over the beam profile
\begin{equation}
    \label{eq:signal}
    N_1(\theta, \chi) = 2\pi \rho \int{\cal R} d{\cal R}\int  d{\cal Z}
    p_1({\cal R,Z}, \theta, \chi ),
\end{equation}
where $\rho$ is the constant density of the target gas. In
experiments it is difficult to measure the absolute yield due to
unknown detection efficiency. The measured ion signal is, however,
proportional to the number of ionized molecules. Measured ratios
of yields for different molecular orientations are thus
independent of detection efficiency. We note that the actual
evaluation of the integral Eq. (\ref{eqn:rate_sol}) can be
simplified by integrating over isointensity shells
\cite{fittinghoff93}. This method describes experiments in which
all ionized molecules are measured.

\section{Results and discussion}
\label{sec:results}

To justify the use of the approximate theories discussed
in Sec.~\ref{sec:theory},
we first calculate the intensity dependent ion signal for
ethylene and benzene and compare with experimental data.
Previous studies using the velocity gauge MO-SFA have shown that
the absence of the final state Coulomb interaction in this model will
underestimate the ionization rates. To take this interaction into account a
correction factor $C_\text{Coul}^{(v)} = (\kappa^3/F_0)^{2
  Z_\text{ion}/\kappa}$ was proposed \cite{muthbohm00}. In our velocity
gauge calculations, we adopt this correction factor. A recent
study on N$_2$ \cite{kjeldsen04} revealed that the effect of the
Coulomb interaction is much smaller with the length gauge MO-SFA.
The explanation is that the combined laser-ion interaction with
the electron is already present through the factor
$\tilde{\Phi}_0(\bm q_n + \bm A(t))$ in Eq.~(\ref{eq:ltransamp})
and also in the length gauge, the ionization occurs at large
distances from the core. In this spatial region, the Coulomb
interaction is already suppressed compared to the electron--laser
interaction. Below we demonstrate that the length gauge MO-SFA
without any extra Coulomb correction factor reproduces
experimental ion signals of C$_2$H$_4$ and
C$_6$H$_6$.

For direct comparison of results obtained by MO-SFA and MO-ADK
theories with the experimental data it is necessary to express the
total rate of ionization as the sum of rates leading to unresolved
final states. This means that the rates in
Eqs.~(\ref{eq:angdiffrate}) and (\ref{eq:totrate}) must be
multiplied by the number of electrons which occupy the active
orbital, i.e., 4 for the degenerate HOMO of benzene and 2 for the
non--degenerate closed shell orbitals.

\begin{figure}
 \centering
\includegraphics[scale=0.4]{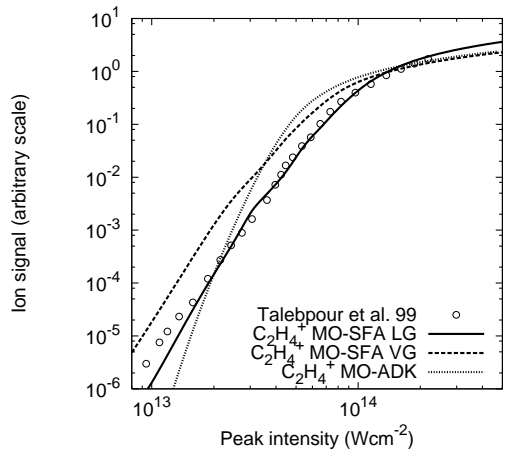}
 \caption{Ion signal for ethylene at 800 nm versus the peak
 intensity of the laser. The laser wavelength is $800\, \mbox{nm}$,
 the beam waist is $49\, \mu \mbox{m}$ and the pulse duration (FWHM)
 is $200\, \mbox{fs}$. Experimental data are from Ref.~\cite{talebpour99}.
 The solid line indicates the the length gauge MO-SFA (MO-SFA LG) predictions,
 the long--dashed line shows the velocity gauge MO-SFA (MO-SFA VG) predictions
 and the short--dashed line shows predictions of MO-ADK theory.
 }
 \label{fig:signal-I-c2h4}
\end{figure}
\begin{figure}
 \centering
\includegraphics[scale=0.4]{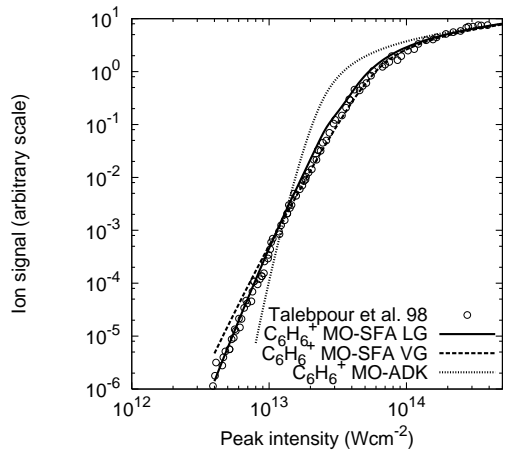}
\caption{Similar to Fig.~\ref{fig:signal-I-c2h4} but for benzene.
Experimental data are from Ref.~\cite{talebpour98}.}
 \label{fig:signal-I-c6h6}
\end{figure}
Figures \ref{fig:signal-I-c2h4} and \ref{fig:signal-I-c6h6} show
theoretical results and experimental data of the ion signal versus
peak intensity for experiments on ethylene and benzene,
respectively. The calculations are the MO-ADK, the length gauge
MO-SFA (MO-SFA LG) and the velocity gauge MO-SFA (MO-SFA VG).
Following the discussion in Sec.~\ref{sec:signal}, we normalize
the results of each calculation to match the experimental data at
high intensities. The figures show that the MO-SFA LG predicts
results in excellent agreement with both the ethylene and the
benzene experimental data. The Coulomb corrected MO-SFA VG results
fit the data reasonably well, while we find a poor agreement
between the MO-ADK model and experiments. We explain this latter
discrepancy by the fact that the intensities considered here do
not strictly correspond to the tunneling regime. Typical values of
the Keldysh parameter are $3.0$ ($2.8$) for ethylene (benzene) at
$I_0 = 10^{13}\, \mbox{Wcm}^{-2}$ and $0.94$ ($0.88$) at $I_0 =
10^{14}\, \mbox{Wcm}^{-2}$. We note that despite both MO-SFA
models are in agreement with these total ion-yield experiments, a
recent study on orientational-dependent ionization from N$_2$
showed that the length gauge formulation is the better
choice~\cite{kjeldsen04}.

\subsection{Ethylene}
\label{sec:ethylene}

The main conclusion from our earlier velocity gauge work on ethylene
\cite{kjeldsen03}
was that the angular distributions of photoelectrons are very sensitive to
the relative orientation between the laser polarization and the molecule.
This could be interpreted as an effect of the nodal plane structure of the
HOMO. Electron emission was found to be impossible in the directions of the
molecular plane which led to a suppression of the total ionization rate
when the molecular plane coincided with the polarization vector.

\begin{figure}
    \begin{center}
        \includegraphics[scale=0.4]{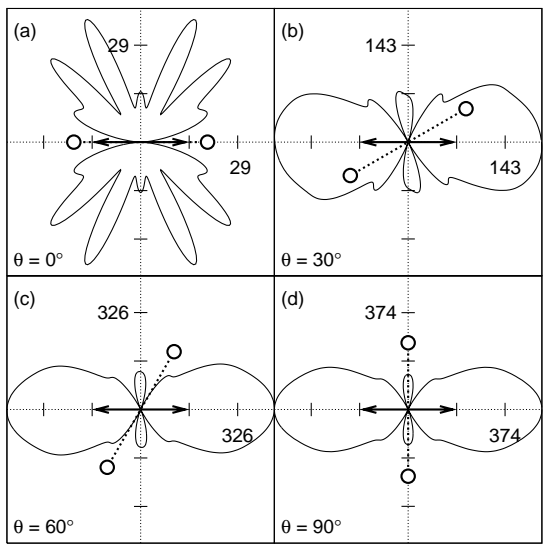}
    \end{center}
    \caption{(a)--(d) Polar plots of the angular differential ionization
    rate of ethylene by an $800$ nm,
    $5 \times 10^{13}\, \mbox{Wcm}^{-2}$ field using the length gauge MO-SFA.  Rates are in units of
      $10^{10}\, \mbox{s}^{-1}$.  The polarization (double
      headed arrow) is horizontal and the molecular plane is perpendicular
      to the paper with the C--C axis (dotted line) in the plane of
      the paper ($\chi$ = 0$^\circ$) and (a) $\theta$ = 0$^{\circ}$,
      (b) 30$^{\circ}$, (c) 60$^{\circ}$, (d) 90$^{\circ}$.
    }
      \label{fig:polplot-len-c2h4-800}
\end{figure}

\begin{figure}
    \begin{center}
        \includegraphics[scale=0.4]{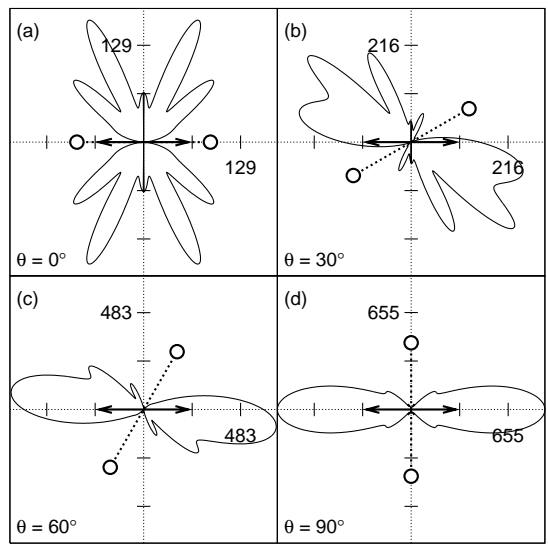}
    \end{center}
    \caption{ Similar to Fig.~\ref{fig:polplot-len-c2h4-800} but calculated
    in velocity gauge.
    }
      \label{fig:polplot-vel-c2h4-800}
\end{figure}

We first present polar plots of angular differential ionization rates using
the length gauge MO-SFA (Fig.~\ref{fig:polplot-len-c2h4-800}) and the Coulomb
corrected velocity gauge MO-SFA (Fig.~\ref{fig:polplot-vel-c2h4-800}) for different
molecular orientations, specified by the angle $\theta$ between the
polarization and the body--fixed $z$ axis defined in
Fig.~\ref{fig:structure}~(a).
Our primary purpose of these calculations is to
check if the results in the two gauges are compatible.
From Figs.~\ref{fig:polplot-len-c2h4-800} and
\ref{fig:polplot-vel-c2h4-800} we see that both models predict an
increase in the ionization rate as the polar angle $\theta$
increases. When the molecular plane coincides with the
polarization, the rates are strongly suppressed and in both models
no electron emission is observed along the polarization direction.
In the velocity gauge this phenomenon was explained by the
presence of a nodal plane in the molecular plane
\cite{kjeldsen03}. There is a one--to--one correspondence between
nodal planes in coordinate space and momentum space. Thus, the
velocity gauge transition amplitude, Eq.~(\ref{eq:vtransamp}),
vanishes at momenta which lie in the nodal plane. This is indeed
what we observe in Fig.~\ref{fig:polplot-vel-c2h4-800}. In the
length gauge transition amplitude, Eq.~(\ref{eq:ltransamp}), the
momentum wave function is not evaluated at the actual outgoing
momentum but at a time--dependent displaced momentum $\bm q_n +
\bm A(t)$. This has the consequence that even though $\bm q_n$
lies in the nodal plane the transition amplitude can be nonzero.
Only when $\bm q_n$ and $\bm A(t)$ both lie in the nodal plane
($\theta = 0^\circ$) the transition amplitude is zero. This
argument is general for molecules which ionize from orbitals
containing nodal planes. In both gauges the ionization rate is
thus expected to be suppressed for molecular orientations with the
polarization coinciding with a nodal plane. This leads to the
qualitative agreement in the two gauges. For molecules without
planar nodal surfaces no such suppressions can {\it a priori} be
expected for any geometry. Based on previous work on strong--field
ionization of N$_2$ \cite{kjeldsen04} and detachment of negative
ions \cite{Gribakin97,beiser04} we judge that the length gauge
is the proper choice of gauge to use in the strong--field
approximation. Accordingly, in the calculations below we will only
use the length gauge MO-SFA.

Currently, angle resolved photoelectron measurements on aligned molecules
have not been performed. However, the first measurements of total ion yields on
aligned molecules have been reported \cite{corkum03}.
\begin{figure}
    \begin{center}
        \includegraphics[width=\columnwidth]{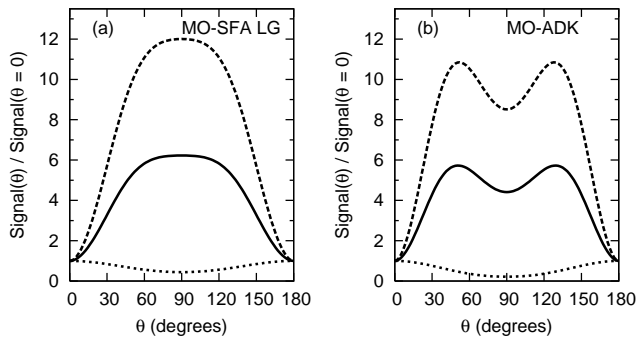}
    \end{center}
    \caption{The orientational dependent ion signal for ionization from
    ethylene using (a) the molecular strong-field approximation in the
    length gauge formulation and (b)
    the molecular
    tunneling theory. The signal is given as a function of the polar
    angle $\theta$ with $\chi = 0^\circ$ (long dashed), $\chi = 90^\circ$
    (short dashed) and $\chi$ averaged (solid).
    The parameters of the laser are:
    $\lambda = 800\, \mbox{nm}$,
    $I_0 = 5 \times 10^{13}\, \mbox{Wcm}^{-2}$, $\tau = 20\, \mbox{fs}$ and
    $w_0 = 49\, \mu\mbox{m}$. }
    \label{fig:sig-geo-c2h4-5E13-800}
\end{figure}
In order to make prediction on these
types of experiments we show in Fig.~\ref{fig:sig-geo-c2h4-5E13-800} the
total ion yield as a function of molecular orientation using (a) the MO-SFA
LG model and (b) the MO-ADK model. The figure shows the
yields which would be measured if all molecules were oriented according to
the Euler angles specified. Explicitly, we show the signals as a
function of the polar angle $\theta$ for two specific $\chi$ angles with all
signals being normalized to the $\theta = 0^\circ$ geometry.
We recall that $\theta = 0^\circ$ for the molecular plane parallel with the
linear polarization, and $\theta = 90^\circ$ for the perpendicular geometry.
In experiments, utilizing linearly
polarized light pulses for the alignment, the molecular axis with
largest polarizability
is aligned along the polarization direction \cite{stapelfeldt03}
(the C--C axis for ethylene).
In this case the molecule is free to rotate
around its body--fixed $z$ axis and measured signals are thus a result of an
average over the $\chi$ angle. Correspondingly, we have calculated the
$\chi$ averaged signal too.

Figure~\ref{fig:sig-geo-c2h4-5E13-800}~(a) shows, in accordance with
Fig.~\ref{fig:polplot-len-c2h4-800}, a large increase in the
ion yield with increasing polar angle for $\chi = 0^\circ$ as well as for
the $\chi$ averaged result. For $\chi = 90^\circ$ the polarization vector is always
directed in the molecular plane and
by the argument given above the yield is generally much lower.
In Fig.~\ref{fig:sig-geo-c2h4-5E13-800}~(b) we see much the same trend for the
MO-ADK model with the $\chi = 90^\circ$ geometries being suppressed compared
with their $\chi = 0^\circ$ counterparts. Contrary to the results from the
MO-SFA LG model, the yield is not maximized when polarization is perpendicular
to the plane for $\chi = 0^\circ$ and $\chi$ averaged.
The theories disagree because the tunneling rate is derived
under the assumption that the electron emission only occurs close to the
space--fixed $Z$ axis determined by the linear polarization direction of the
field. Hence, the MO-ADK rates reflect only the situation in
this spatial region. For atoms, this is a reasonable assumption since the initial
spherical symmetric potential will be affected most strongly in this
direction and the potential barrier which arises from the combined field and
atomic potential will then be minimized. For molecules,
the initial potential is more complicated and the potential barrier can be
minimized in other directions. Important contributions to the ion yield are
thereby not taken into account and this is the reason why the results of the theories do
not agree.

\begin{figure}
    \begin{center}
        \includegraphics[width=\columnwidth]{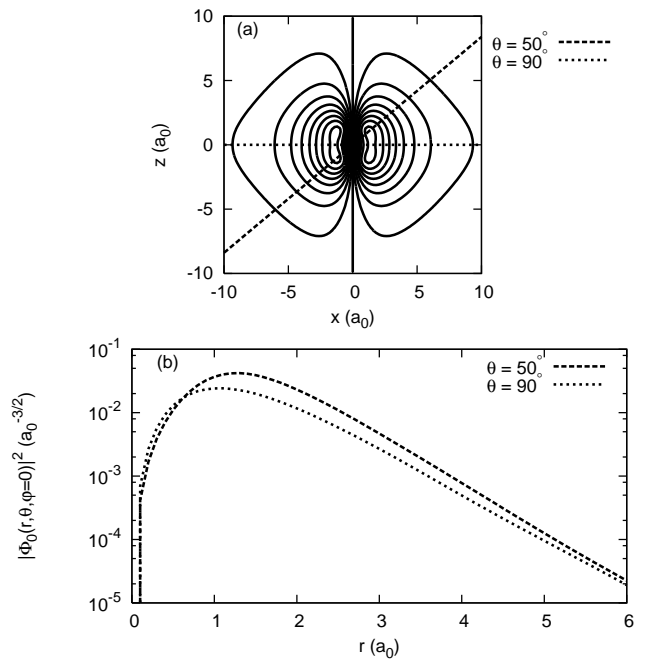}
    \end{center}
    \caption{(a) Contour plot of the HOMO of ethylene in the $xz$ plane. (b)
    The norm square of the HOMO along the directions $\theta = 50^\circ$ and
    $\theta = 90^\circ$. The unit of length is the Bohr radius $a_0 = 5.29177
    \times 10^{-11}\, \mbox{m}$.}
    \label{fig:contour}
\end{figure}

Since the tunneling ionization occurs close to the $Z$ axis by
assumption, the interpretation of the MO-ADK results can be made
by analyzing the electron distribution in this direction.
Previously, it was noted that the tunneling rate is largest when
the initial electronic cloud is aligned with the field direction
\cite{lin02}. In Fig.~\ref{fig:contour}~(a) we show a contour plot
of the HOMO in the body fixed $xz$ plane. When the field vector
lies in this plane the geometry corresponds to $\chi = 0^\circ$
and with vertical and horizontal directions corresponding to
$\theta = 0^\circ$ and $\theta = 90^\circ$, respectively. To make
it clear that $\theta = 90^\circ$ does not correspond to the
largest electron density we have furthermore made a plot of
$|\Phi_0(\bm r)|^2$ along the directions $\theta = 50^\circ$ and
$\theta = 90^\circ$ (Fig.~\ref{fig:contour}~(b)). We clearly see
that the electron density is largest for $\theta = 50^\circ$ at
large distances.
The minimum around  $\theta = 90^\circ$ is also present in
Fig.~\ref{fig:sig-geo-c2h4-5E13-800} (b) for the $\chi$
averaged yields.

\subsection{Benzene}
\label{sec:benzene}

\begin{figure}
    \begin{center}
        \includegraphics[width=\columnwidth]{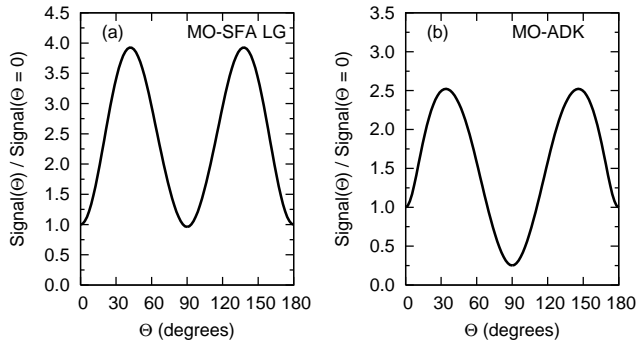}
    \end{center}
    \caption{The orientational dependent ion signal for ionization of
    benzene as a function of $\Theta = 90^\circ - \theta$.
    The parameters of the laser are given in the caption of
    Fig.~\ref{fig:sig-geo-c2h4-5E13-800}. }
    \label{fig:sig-geo-c6h6-5E13-800}
\end{figure}

Benzene is a symmetric top molecule and hence it can only be aligned in the
molecular plane \cite{stapelfeldt03} while it will be free to rotate around
the body--fixed $z$ axis (see Fig.~\ref{fig:structure} (b)). 
Thus only $\chi$ averaged signals as shown in
Fig.~\ref{fig:sig-geo-c6h6-5E13-800} can be measured.
We choose to present our calculated signals as a function of the
angle, $\Theta$, between the molecular plane and the polarization. Note that
this angle is related to the usual Euler angle  through the
relation $\Theta = 90^\circ - \theta$ where $\theta$ is the polar angle
between the $Z$--axis and the moleular plane in the frame defined in
Fig.~\ref{fig:structure}. This choice of data representation is
convenient since a comparison with benzene derivatives like
fluorobenzene and chlorofluorobenzene can be made more directly
(Secs.~\ref{sec:fluorobenzene} and \ref{sec:chlorofluorobenzene}).
In Fig.~\ref{fig:sig-geo-c6h6-5E13-800} we see that the MO--SFA LG model
predicts the maximum yields at $\Theta = 42^\circ$ and $\Theta = 138^\circ$
while the MO--ADK model predicts maximum yields at $\Theta = 33^\circ$ and
$\Theta = 147^\circ$. In both models the yield is minimized at
$\Theta = 0^\circ$, $90^\circ$ and $180^\circ$.
These findings are in accordance with our expectations
\cite{kjeldsen03} since for $\Theta = 0^\circ$, $90^\circ$ and $180^\circ$ 
the polarization vector lies in a nodal plane and  the
asymptotic electron density is maximized in between. In a
quantitative comparison between the models we find a somewhat
larger modulation in the orientational dependent yield with the
MO-ADK model than in the MO-SFA LG model. 

\subsection{Fluorobenzene}
\label{sec:fluorobenzene}

When one hydrogen atom in benzene is replaced by a fluorine atom, the
six--fold rotational symmetry is broken and consequently the
degeneracy of the HOMO is lifted. The doubly degenerate HOMO of benzene splits up in two
components with the ionization potentials $9.35\, \mbox{eV}$
and $9.75\, \mbox{eV}$.
This energy difference is small compared with the energy separations
to the other occupied orbitals and we will thus consider
ionization from both orbitals.

\begin{figure}
    \begin{center}
      \includegraphics[width=\columnwidth]{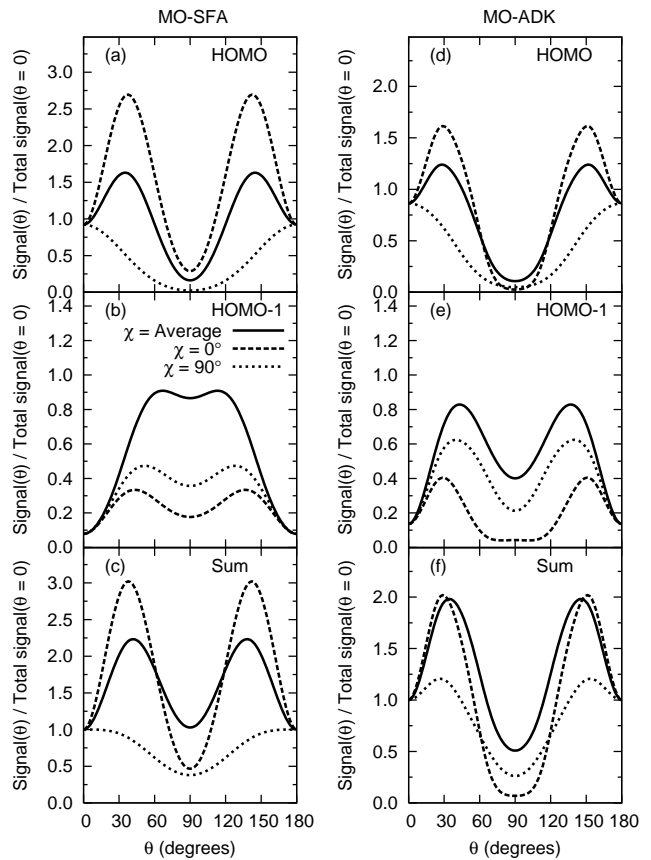}
    \end{center}
    \caption{
    The orientational dependent ion signal for ionization from
    the highest occupied orbitals of fluorobenzene using MO-SFA LG (a)-(c) and
    MO-ADK (d)-(f).
    See Fig.~\ref{fig:sig-geo-c2h4-5E13-800} for details on the laser
    pulse and legends.}
    \label{fig:sig-geo-c6h5f-5E13-800}
\end{figure}
Using the MO-SFA LG model, we present in
Fig.~\ref{fig:sig-geo-c6h5f-5E13-800} (a)--(c) the ion yields
which originate from (a) the HOMO, (b) the second highest occupied
orbital (\mbox{HOMO--1}) and (c) the integral yield from both
channels. The figures \ref{fig:sig-geo-c6h5f-5E13-800}~(d)--(f)
contain equivalent results from the MO-ADK model. We find that the
signals from both orbitals contribute significantly to the total
signal and depending on the geometry, the contribution from either
one will dominate. The relative importance between the ionization
channels at a given geometry can be understood as a consequence of
the symmetries of the orbitals.

Due to the $C_{2v}$ symmetry of fluorobenzene the orbitals are eigenstates
of the $xz$ reflection operator with eigenvalues $+1$ (even) or $-1$ (odd).
The reflection symmetry of an orbital is indicated by the subscript 1
for the even states and 2 for the odd states. The HOMO of
fluorobenzene is a $b_1$ orbital, Fig.~\ref{fig:structure}~(c),
while the HOMO--1 is an $a_2$ orbital, Fig.~\ref{fig:structure}~(d).
Both orbitals are odd with respect to
reflections in the molecular ($yz$) plane. Similar to ethylene, ionization
of fluorobenzene is suppressed when $\chi = 90^\circ$ compared with the
$\chi$ average since the molecular plane is a nodal plane for both
active orbitals. The $b_1$ orbital contains a non--planar nodal surface
which is approximately parallel to the $xy$ plane. When $\theta = 90^\circ$
the polarization is accordingly nearly parallel to this surface and we
find the yield from the HOMO to be minimized around $\theta = 90^\circ$.
The $a_2$ orbital is odd under the $xz$ reflection and thus it contains a nodal
plane in the $xz$ plane. When the polarization is parallel to this
plane ($\chi = 0^\circ $) we also find suppression of the ionization signal.
In all cases we find qualitative agreement between the MO-SFA LG and MO-ADK
results. In order to discuss the effect of the substitution
of hydrogen with fluorine we note that the angles $\Theta$ for benzene
and $\theta$ with
$\chi = 0^\circ$ for fluorobenzene both describe the angle between the respective
molecular planes and the polarization. It is thus relevant to compare the
curve in Fig.~\ref{fig:sig-geo-c6h6-5E13-800} with the dashed curve in
Fig.~\ref{fig:sig-geo-c6h5f-5E13-800}.
In both theories we find that the total yield
for fluorobenzene is slightly more suppressed around $\theta = 90^\circ$ than for benzene
when it is compared with the maximum yields around 45 degrees (MO-SFA LG)
and 30 degrees (MO-ADK).

\subsection{Chlorofluorobenzene}
\label{sec:chlorofluorobenzene}

\begin{figure}
    \begin{center}
        \includegraphics[width=\columnwidth]{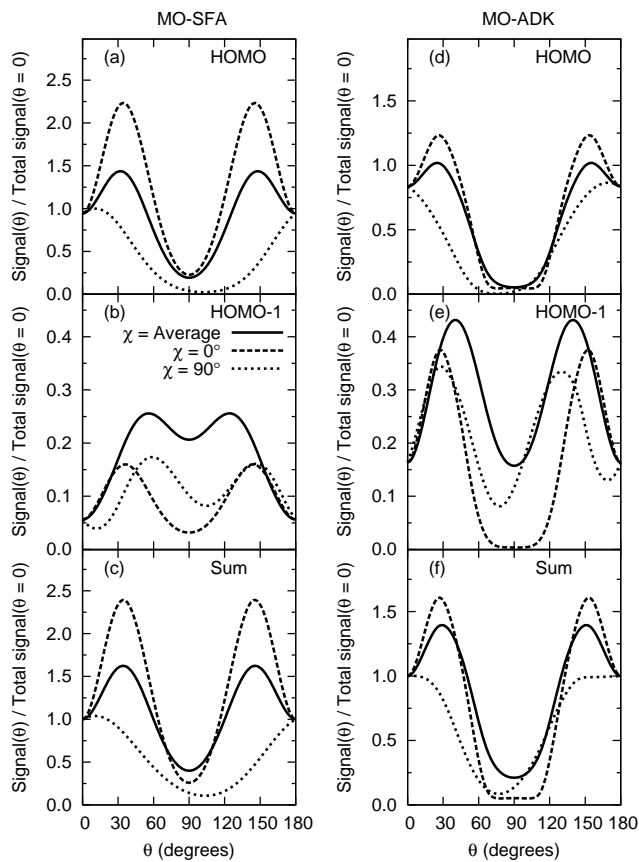}
    \end{center}
    \caption{
    Similar to Fig.~\ref{fig:sig-geo-c6h5f-5E13-800} but for
    chlorofluorobenzene.
    }
    \label{fig:sig-geo-c6h4clf-5E13-800}
\end{figure}
In Fig.~\ref{fig:sig-geo-c6h4clf-5E13-800} we present the series of graphs
for chlorofluorobenzene equivalent to the fluorobenzene results of
Fig.~\ref{fig:sig-geo-c6h5f-5E13-800}.
The ionization potentials that correspond to the two highest occupied
orbitals are $9.37\, \mbox{eV}$ and $9.80\, \mbox{eV}$, meaning that the
energy difference between the HOMO and the HOMO--1 is  $0.43\, \mbox{eV}$
\cite{mohraz80}.

The molecular symmetry
is reduced to contain only the reflection in the molecular plane as a symmetry
operation ($C_s$ point group) and, consequently, planar nodal surfaces can only lie in this
plane. However, the
two active orbitals of chlorofluorobenzene bear some similarities with their
counterparts in fluorobenzene and thus our qualitative predictions will also
be similar.
The similar nodal surface structures can be
seen by comparison of Figs.~\ref{fig:structure}~(c)--(d) with
Figs.~\ref{fig:structure}~(e)--(f). 
The HOMO contains a nodal surface approximately parallel to the $xy$ plane
and the HOMO--1 contains a nodal surface approximately parallel to the
$xz$ plane. The similarities between the orbitals can be further exploited by
comparing the asymptotic coefficients which are given in table
\ref{tab:clm}.

When we
compare Figs.~\ref{fig:sig-geo-c6h5f-5E13-800} and
\ref{fig:sig-geo-c6h4clf-5E13-800} we do indeed find alike behaviour. For the second
highest orbital the suppression at $\chi = 0^\circ$ is now less pronounced. We can explain
this phenomenon by the fact that the polarization is not parallel to a nodal
plane but only approximately parallel to a nodal surface.

The most significant difference between the MO-SFA LG and the
MO-ADK model is found at $\chi = 90^\circ$. Here the molecular
plane lies parallel to the polarization and the angle $\theta$
determines how the molecule is rotated in this plane. In the
MO-SFA LG model the total yield is minimized when $\theta >
90^\circ$ which corresponds to a geometry where the chlorine atom
is directed along the polarization vector. In the MO-ADK model the
minimum yield is found at $\theta < 90^\circ$ where the fluorine
atom is directed along the polarization vector. We expect that this
discrepancy arises from the fact that the tunneling theory only
accounts for ionization in a narrow cylinder along the
polarization direction of the field, as it is discussed in
Sec.~\ref{sec:ethylene}, and thus the MO-SFA LG model predicts the
more accurate results.

\section{Conclusions and outlook}
\label{sec:conclusions}

In the present work, we have given a detailed account of the
MO-SFA in the length gauge formulation, and we have presented the
MO-ADK tunneling theory for non-linear polyatomic molecules. We
have discussed the validity of the present effective
single-electron models, and we have applied them to molecular
systems where we expect the approach to be well justified. Indeed
a comparison of our theoretical results with experimental
ion-signal data for ethylene and benzene showed very good
agreement, in particular in the length gauge formulation of the
molecular strong-field approximation.

We have presented a detailed study of the characteristics of
orientational-dependent ionization signals for ethylene, benzene,
fluorobenzene, and chlorofluorobenzene, and thus covered systems
with $D_{2h}$, $D_{6h}$, $C_{2v}$, and $ C_s$ point group
symmetries. Our calculations of angular differential rates for
ethylene showed that the distributions are largely determined by
the nodal plane structure, and the comparison of predictions of the
MO-ADK and the MO-SFA results for the orientational dependent ion signal
for ethylene illustrated the shortcoming of the former model in
accounting for situations where the ionization is peaked in other
directions than along the polarization direction. The comparison
of the orientational-dependent ion signals for the four different
molecules showed that it is in general the nodal plane surfaces
that determine the structure in the ionization yields. The point
group of the molecules, although related,  plays a less important
role. An interesting feature, however, associated with the
reduction in the point group symmetry is the lift of the
degeneracy in the HOMO as we go from benzene to the
substituted benzene molecules. In fluorobenzene and
chlorofluorobenzene the energy difference between the HOMO and the
HOMO--1 is comparably small, and one needs to consider ionization
from both orbitals. The conclusions for the HOMO--1 with respect to
the importance of nodal surfaces remain unchanged. If one, on the
other hand, speculates about possible experimental studies aiming
at the separation and unique identification of HOMO and HOMO--1
dynamics, respectively, we propose to turn to a study of the
photoelectron spectrum or above-threshold-ionization (ATI)
spectrum. In the fixed--nuclei approximation applied throughout
this work, we would see an ATI spectrum consisting of two series
corresponding to electron energies which fulfil $q_n^2/2 = n
\omega - E_b(j) - U_p$ with $E_b(j)$, $(j=\{ {\rm HOMO} , \textrm{HOMO--1}\})$
denoting the binding energies of the orbitals. Our present results
show that depending on the molecular orientation, either one of
the series will lead to the stronger peaks. This effect is of
course absent in benzene. Beyond the fixed--nuclei approximation,
the ATI peaks will broaden due to coupling with the rovibrational
degrees of freedom, and the different series may be impossible to
distinguish. Whether this happens will be a study for the future.

\begin{acknowledgments}
  LBM is supported by the Danish Natural Science Research Council (Grant
No. 21-03-0163).
\end{acknowledgments}

\end{document}